\documentstyle[aps,prl,floats,graphicx,amssymb]{revtex}
\begin{document}
\draft
\twocolumn[\hsize\textwidth\columnwidth\hsize\csname @twocolumnfalse\endcsname
%
  \title{Coexistence of Single and Double-Quantum Vortex Lines}

\author{\"U. Parts$^a$, V.V. Avilov$^b$, J.H. Koivuniemi$^a$, N.B.
Kopnin$^{a,c}$,
M. Krusius$^a$, J.J. Ruohio$^a$, and V.M.H. Ruutu$^a$}
\address{$^a$Low Temperature Laboratory, Helsinki University of
    Technology, P.O.Box 2200, FIN--02015 HUT, Finland\\
    $^b$Department of Magnetohydrodynamics,
    Research Center Rossendorf, P.O.Box 510119, D--01314 Dresden,
    Germany \\
    $^c$ L. D. Landau Institute for Theoretical Physics, 117334 Moscow,
    Russia }
\date{\today}
\maketitle
\begin{abstract}
We discuss the configurations in which singly and doubly quantized
vortex lines may coexist in a rotating superfluid. General
principles of energy minimization lead to the conclusion that in
equilibrium the two vortex species segregate within a cylindrical
vortex cluster in two coaxial domains where the singly quantized
lines are in the outer annular region. This is confirmed with
simulation calculations on discrete vortex lines. Experimentally
the coexistence can be studied in rotating superfluid $^3$He-A.
With cw NMR techniques we find the radial distribution of the two
vortex species to depend on how the cluster is prepared: (i) By
cooling through $T_c$ in rotation, coexistence in the minimum
energy configuration is confirmed. (ii) A glassy agglomerate is
formed if one starts with an equilibrium cluster of single-quantum
vortex lines and adds to it sequentially double-quantum lines, by
increasing the rotation velocity in the superfluid state. This
proves that the energy barriers, which separate different cluster
configurations, are too high for metastabilities to anneal.

\end{abstract}
\pacs{PACS:  67.57.Fg, 52.25.Wz, 74.60.Ge}
\bigskip
] 
\section{Introduction}
Superfluid $^3$He-A is the only known quantum system where
topologically stable vortex lines with different quantization
appear simultaneously.  The doubly quantized vortex (DQV) is
formed at low critical velocity and is usually obtained when
superfluid $^3$He-A is accelerated to motion, eg. in a rotating
container \cite{Ruutu1}. The singly quantized  vortex (SQV), in
turn, may have lower energy, depending on the magnitude of the
magnetic field and flow velocity, but because of its large
critical velocity it is in practice only formed on cooling through
the superfluid transition in nonzero flow \cite{Parts1}.

Normally one will find either double or single-quantum vortex
lines in a rotating container, depending on how the state has been
prepared. In the homogeneous rotating superfluid, transitions
between different  vortex textures  are of first order and
therefore in principle sharp. However, topological stability
forbids a transition from one type of existing vortex lines to
another. Thus SQV and DQV lines, once they have been formed, will
maintain their quantization in the rotating container, as long as
they are not removed by deceleration and annihilation on the
container wall or by warming up to the normal phase. This
metastability can be exploited to prepare a rotating state where
both types of vortex lines are simultaneously present.

{\it A priori} it is not clear what configuration a rotating state
with two different vortex species would take: Do the arrays always
have the same configuration or does this depend on the process by
which the state has been created? Similar questions have been
studied in the case of  point-like objects, namely trapped ions in
a 2 or 3-dimensional point charge system \cite{Dubin}. We
investigate the coexistence of SQV and DQV lines both
theoretically and experimentally. We study the free energies of
different vortex-array configurations, which consist of a mixture
of the two species or of segregated domains. We apply both the
continuum model and use numerical simulation of discrete vortex
lines to show that a configuration with SQV vortices surrounding
an inner central cluster of DQV is lower in energy than any other
configuration. The energy difference is small and disappears in
the continuum approximation.

Axially symmetric  configurations of coexisting SQV and DQV lines
are observed in our measurements when liquid $^3$He is slowly
cooled at constant rotation from the normal phase to $^3$He-A. Our
observations of these states agree with the theoretical
predictions: the SQV lines are found in an outer annulus which
surrounds a central domain of DQV lines. Segregated
configurations, which do not have full axial rotation symmetry,
can also be prepared. This is accomplished by adding DQV lines to
an existing equilibrium SQV cluster, by slowly increasing the
rotation velocity at constant temperature in the superfluid state.
The resulting configuration does not correspond to the global
energy minimum, but rather resembles a glassy agglomerate.

\section{ Rotating $^3$He-A.}

Rotating vortex states are driven and stabilized by superfluid
counterflow (CF), the difference in the velocities of the normal
and superfluid fractions, ${\bf v} = {\bf v}_n - {\bf v}_s$. The
equilibrium states are found by minimizing the free energy
functional \cite{Wolfle}, which controls the spatial distribution
of the order parameter. On the macroscopic scale,  vortex textures
in $^3$He-A are rectilinear structures, oriented parallel to the
rotation axis, which extend from the top to the bottom of the
rotating container. These vortex cells are confined by the Magnus
force to a regular array, a cluster of vortex lines. The cluster
is coaxial with the cylinder \cite{VorBundle} and isolated from
the cylindrical wall by an annular layer of vortex-free CF. Within
the cluster the areal density of linear vortex cells  has the
solid-body-rotation value $n = 2\Omega/\kappa$. Here  $\kappa =
\nu \kappa_0$ is the circulation of a vortex line with the quantum
number $\nu$ and the circulation quantum $\kappa_0 = h/(2m_3)=
0.066$ mm$^2$/s.

In a vortex cluster of infinite diameter a transverse cross
section would display a perfect 2-dimensional lattice with
triangular nearest-neighbor coordination, which arises from the
inter-vortex repulsion. In a finite cluster the central region is
expected to display a triangular lattice, which approaches
crystalline order, but towards the outer boundary it is deformed
by the surrounding annular CF more and more towards a
configuration of concentric rings of vortex lines \cite{Campbell}.
Due to the competition between volume and surface interactions
long range order is thus relatively poor in finite-size arrays at
the usual experimentally accessible rotation velocities. Here the
total number of lines is typically a few hundred, the radius $R$
of the container a few mm, the radius of the Wigner--Seitz (WZ)
cell of the vortex lattice $r_0 \sim 100 \; \mu$m, and the vortex
core radius $\xi \sim 10\; \mu$m \cite{SoftCore}.

The order-parameter texture within a unit cell of the periodic
vortex lattice has been derived from numerical minimization of the
free energy as a function of the rotation velocity ${\bf \Omega}$
and the applied field ${\bf H}$ \cite{Volovik1,Parts1,Thuneberg}.
These calculations reveal a complex phase diagram
\cite{Parts1,Thuneberg}, with quantized vorticity organized in
several topologically and structurally different vortex textures
\cite{Reviews}. SQV vortex textures, which include a singular
vortex core, have only been described approximately, by solving
for the order parameter distribution with analytical trial
functions \cite{Volovik2}. Of these different vortex textures we
shall here only be concerned with two, the continuous DQV and the
singular SQV.

In an array consisting of DQV lines the inter-vortex distance is
larger by $\sqrt{2}$ than in one formed from SQV lines. With
coexisting SQV and DQV lines this large difference opens the
question whether sufficient mismatch and distortion is created in
the nearest-neighbor coordination that a disordered glassy state
results. Or could it perhaps lead to a ``chemical'' structure
where both species occupy fixed positions within each unit cell of
a periodic vortex lattice.  A third possibility is an array
consisting of segregated domains of one species only. These are
the questions which we shall discuss next.

\section{ Energy minimization}

\subsection{Continuum approximation}

On the macroscopic level the energy difference between SQV and DQV
arrays depends on the difference in the quantization numbers and
inter-vortex distances, if the latter are much larger that the
core sizes. Here we neglect the smaller energy contributions
associated with the structure of the  vortex core. Consider the
free energy of the rotating superfluid
\begin{equation}
\tilde{E}\{ {\bf v}_s\}=E_{kin}-L\Omega =\frac{\rho_s}{
2}\int_{r<R} ({\bf v}_s^2-2 {\bf v}_s {\bf v}_{\rm rot})d^2r~~,
 \label{w2.01}
\end{equation}
where ${\bf v}_{\rm rot}=\Omega \, [{\hat{\bf z}\times }{\bf r}]$,
$\Omega$ is the rotation velocity, and $L$ the angular momentum of
the fluid. We express the superflow velocity ${\bf v}_s$ in the
continuum approximation and assume the logarithmic approximation
for the vortex line energy. In the laboratory frame we then have
\cite{Khalat}
\begin{eqnarray}
\tilde E =\frac{\rho _s}{2} \int \left[ v_s^2 + {1 \over {2\pi}}
\, \kappa \,[{\rm curl}\, {\bf v}_s]_z \; \ln \left(\frac{r_0}{\xi
} \right) \right. \nonumber \\ \left. -2\Omega v_s r -\frac{\kappa
}{\pi} \, \Omega \right]\, d^2r \,. \label{energy}
\end{eqnarray}
The second term in the integral represents the energy of a vortex
line. If different vortex species are present, the circulation
$\kappa_a =\nu_a \kappa _0$, the inter-vortex distance $r_a$, and
the core radius $\xi_a$ become functions of the spatial
coordinates (here $a = 1$ or 2).

Let us consider a configuration with $N_1$ SQV lines and
$N_2$ DQV lines. The overall radius $R_v$ of the vortex cluster is
determined by the total circulation:
\begin{equation}
N_1\kappa _1+N_2\kappa _2=2\pi \Omega R_{v}^2 \; . \label{Ncond}
\end{equation}
In the stationary state the areal densities of vortex lines are
proportional to their circulation,
\begin{equation}
\pi r_1^2=\frac{\kappa _1}{2 \Omega};~~\pi r_2^2=\frac{\kappa
_2}{2 \Omega}\;. \label{WS-radius}
\end{equation}
The average velocity field is ${\bf v}_s={\bf v}_{s0}$ where
\begin{equation}
{\bf v}_{s0}=\cases{{\bf v}_{\rm rot}, & if $r<R_v$\cr \Omega \,
[\hat{\bf z}\times {\bf r}] \, R_v^2/r^2 ,& if $ r>R_v$. \cr }
\label{v/contin}
\end{equation}
The free energy in Eq.~(\ref{energy}) is now given by
\begin{equation}
\frac{\tilde E}{E_0}=
\left(\frac{R_{v}}{R}\right)^4\left[\frac{3}{4}- \ln
\left(\frac{R_v}{R}\right)-\left(\frac{R}{R_v}\right)^{2} \right]
+\frac{S_1\epsilon _1+S_2\epsilon _2}{S} \label{energy1/ab}
\end{equation}
where $E_0 = \pi \rho _s \Omega ^2R^4$, $S=\pi R^2$, and
$S_a=\kappa_a N_a/(2\Omega)$ is the area occupied by SQV or DQV
lines, such that $S_1+S_2=\pi R_v^2$. The normalized vortex-line
energy densities in the logarithmic approximation (see
Eq.~(\ref{WS-cylinder})) are
\begin{equation}
\epsilon _1=\frac{r^2_1}{ R^2} \ln \left(\frac{r_1}{\sqrt{e}\xi
_1}\right);~~ \epsilon _2=\frac{r^2_2}{R^2} \ln
\left(\frac{r_2}{\sqrt{e}\xi _2}\right) \; . \label{epsilon}
\end{equation}
These are proportional to the first power of $\kappa _a$ as
distinct from the single-vortex energy which is proportional to
$\kappa _a^2$ (where $a=1$ or 2).

These expressions are valid when $r_a \gg \xi_a$ \cite{SoftCore}.
The energy densities of SQV and DQV  arrays become equal when
\[
\left(\frac{\kappa _1}{2\pi\Omega e \xi_1^2}\right)^{\kappa_1} =
\left(\frac{\kappa _2}{2\pi\Omega e \xi _2^2}\right)^{\kappa_2}
\;.
\]
The rotation velocity for the transition becomes
\begin{equation}
\Omega^* = {{2\kappa _0} \over {\pi e}} \, {\xi_1^2 \over \xi_2^4}
\; . \label{Omega*}
\end{equation}

To explore the consequences from the continuum approximation
further, we minimize the energy in Eq.~(\ref{energy1/ab}) with
respect to variations of both $N_1$ and $N_2$. This gives the
equilibrium configuration, which one would expect to find after an
adiabatic transition at constant rotation from the normal to the
superfluid state. Several conclusions follow:

(1) If $\epsilon _2>\epsilon _1$, only SQV lines remain since
$S_2$ shrinks to zero (Eq.~(\ref{energy1/ab})).  In the opposite
case, $S_1$ shrinks to zero, and only DQV lines remain. The DQV
becomes more favorable, when $\Omega >\Omega^*$. According to our
measurements this occurs above 0.6 rad/s
(Sec.~\ref{Sec:RotDependence}).

(2) The radius $R_v$ of the cluster has its equilibrium value when
the width of the surrounding vortex-free CF annulus becomes
\begin{equation}
d_{a(eq)}= R\sqrt{\epsilon _a/2}\;. \label{e:deq}
\end{equation}
where the outer boundary of the cluster consists of type $a$
vortices.

(3) The energy in Eq.~(\ref{energy1/ab}) does not depend on the
configuration of the vortex cluster with given $N_1$ and $N_2$.
Consequently, the continuum approximation in the logarithmic limit
does not discriminate between different vortex cluster
configurations.

\subsection{ Discrete vortex lines}
\subsubsection{Sequentially formed arrays}

The continuum approximation in Eq.~(\ref{energy1/ab}) does not
include the surface energies at the interfaces between the vortex
cluster and the vortex-free region or between segregated domains
occupied by different vortex species. Nor does it account for
interactions between individual vortex lines and the container
wall. To identify configuration-dependent contributions to the
free energy, we need to look at vortex clusters composed of
discrete vortex lines. We perform numerical calculations on vortex
arrays which consist of $N_j$ discrete rectilinear vortex lines
with coordinates ${\bf r}_j$, circulations $\kappa _j=\nu_j
\kappa_0$, where $\nu _j=1, 2$, and core radii $\xi (j)$. The
superfluid velocity produced by the lines is
\begin{equation}
{\bf v}_s=\hat{\bf z}\times \sum_{j} {\kappa_j \over 2\pi} \left[
{({\bf r-r}_j )\over\mid {\bf r-r}_j \mid^2}- {({\bf r-r}_j
')\over\mid {\bf r-r}_j ' \mid^2} \right]~~. \label{w0.01a}
\end{equation}
The last term accounts for the image vortices with coordinates
${\bf r}_j^\prime ={\bf r}_j R^2/r_j^2 $, of which each one
corresponds to one true vortex located at ${\bf r}_j$.

The interaction potential for vortex lines in the rotating
superfluid is equivalent to the Coulomb potential between charged
lines immersed into a uniform background charge of opposite sign.
The density of the background charge is proportional to $\Omega$.
The potential acting on each line is the sum of the potentials of
the background charge, of the direct logarithmic Coulomb interline
interaction, and the interactions with the wall reflections.

The equation describing the dynamics of vortex lines has the form
\begin{equation}
\dot{\bf r}_j =\left(1-b^\prime \right){\bf v}_s^\prime
-b[\hat{\bf z}\times {\bf v}_s^\prime ]~, \label{eq.of.motion}
\end{equation}
where $\dot{\bf r}_j$ is the velocity of the $j$th vortex line and
${\bf v}_s^\prime $ the superfluid velocity produced by all other
vortices at the position of the $j$th reference line, while $b$
and $b^\prime$ are the reduced Hall--Vinen mutual friction
parameters \cite{Hook}.

\begin{figure}[!t]
  \begin{center}
    \leavevmode
    \includegraphics[width=0.6\linewidth]{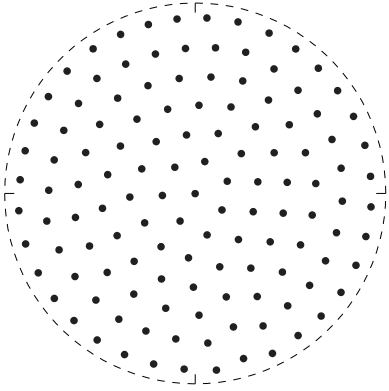}
    \bigskip
    \caption{
Computer simulation of vortex array formation. Here $N_1=127$
rectilinear SQV lines have been accumulated one by one to form a
cluster, by increasing $\Omega$ slowly in the superfluid state.
The equation of motion (\protect\ref{eq.of.motion}) with $b^\prime
=0$ has been used. The nucleation center, from where the lines are
injected, is located at the cylinder wall at $x = R$ (not shown)
and $y=0$.
The final configuration of the array in the cylindrical container
is shown. The dashed circle marks the edge of the cluster at
$r=R_v$. In the center the lines form a 2D hexagonal crystal with
a large number of dislocations, whereas towards the perimeter the
configuration deforms to circular concentric rings. }
  \label{GrowthSQV}
  \end{center}
  \end{figure}

To illustrate the dynamic simulation in vortex array formation, we
consider a standard experiment when a vortex array is formed in
the superfluid state by increasing slowly  the rotation velocity
linearly from zero \cite{Ruutu1}. In this process one vortex line
is formed after another periodically, every time when the CF
velocity at the cylindrical wall exceeds a constant critical
value. Fig.~\ref{GrowthSQV} illustrates a typical final
configuration for such a cluster which includes only one species
of vortex lines. They are introduced at a fixed nucleation center,
once the CF velocity reaches the critical value.

To simplify matters we may here think of the nucleation center in
the form of a line source parallel to the rotation axis, located
close to the cylindrical wall of the container \cite{Ruutu1}. From
the nucleation center the newly created rectilinear vortex line
moves to the edge of the vortex cluster under the action of the
background potential of uniform rotation along a spiral
trajectory, which depends on the values of the mutual friction
parameters $b$ and $b^\prime$. The line is incorporated into the
cluster by pushing via the repulsive interactions, which leads to
plastic deformations in the existing array. The final structure of
the array resembles in the central region a hexagonal lattice with
a large number of dislocations. The lines in the periphery form
circular rings which are separated by a wide transition region
from the central crystal.

\begin{figure}[!t]
  \begin{center}
    \leavevmode
    \includegraphics[width=0.9\linewidth]{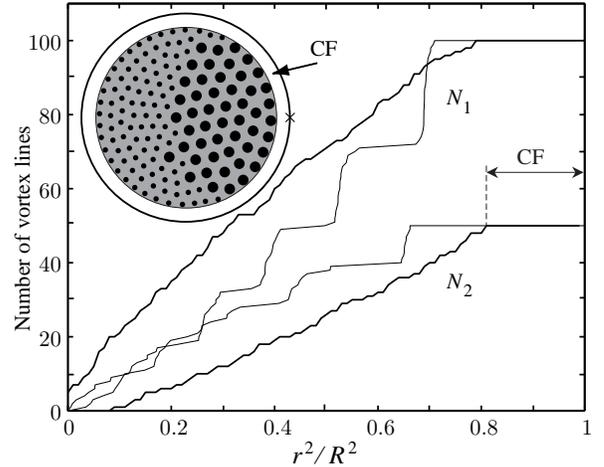}
    \bigskip
    \caption{
Simulation calculation of an array with 100 SQV and 50 DQV lines.
The DQV lines have been accumulated one by one to an existing
equilibrium array of SQV lines, similar to the experiment in
Fig.~\protect\ref{Glass}. The nucleation center, from where the
DQV lines are injected, is located on the cylindrical wall ($x=R,
y=0)$ and is marked with a cross in the {\it inset} where the
final configuration of the array at 0.49 rad/s is shown. With this
number of vortices the array would correspond to equilibrium
circulation at 0.34 rad/s. The thin curves in the {\it main panel}
represent the radial locations of the SQV and DQV lines in the
array: $N_a(r) = \sum_j (x_{a,j}^2 + y_{a,j}^2)$. They illustrate
how the locations of the vortex lines concentrate with increasing
radius more and more on concentric rings, although the
nearest-neighbor coordination is still triangular, as seen in the
inset. The thick curves correspond to deceleration records
($\Omega:$ 0.49 rad/s $\rightarrow 0$), similar to those measured
in the experiment (Fig.~\protect\ref{Glass}), and give the average
radial distributions $N_1(r^2)$ and $N_2(r^2)$. }
  \label{SimulPlot1}
  \end{center}
  \end{figure}

A second variation of the same calculation, but now for a
two-component vortex array, is shown in Fig.~\ref{SimulPlot1}.
Again the final state is shown, after DQV lines have been
accumulated one by one into an existing cluster of SQV lines.
Despite large plastic deformations in the cluster, the injected
DQV vortex lines form a compact drop. We observe neither a
periodic (chemical) or irregular structure of single DQV lines
inside the SQV array. Instead this simulation procedure gives a
glassy vortex agglomerate with hexagonal coordination, which is
consistent with the experimental situation in $^3$He-A (as will be
seen below by comparing eg. with Fig.~\ref{Glass}).

\begin{figure}[!t]
  \begin{center}
    \leavevmode
    \includegraphics[width=0.7\linewidth]{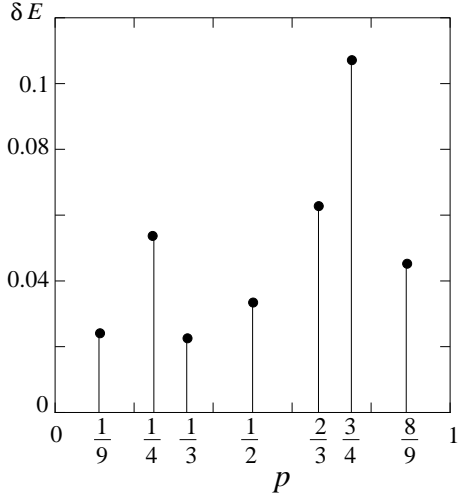}
    \bigskip
    \caption{
Energy of mixing $\delta E$ of different periodic 2D vortex line
lattices, formed from SQV and DQV lines. The zero level
corresponds to the energy of two separated hexagonal SQV and DQV
lattices. The unit cell includes $N_1$ SQV and $N_2$ DQV lines.
The horizontal axis gives the fraction $p=N_2/N_c$, where
$N_c=N_1+N_2$. The energy of mixing $\delta E$ is expressed in
units of $\rho_s \kappa_0^2 /(2\pi N_c)$. The lattices with
$N_c=2$ and $N_c=4$ have a square unit cell, while all other
lattices are obtained by inserting the additional vortex lines
into a hexagonal cell.}
  \label{EMmixture}
  \end{center}
  \end{figure}

To explain these results let us first consider the energy of
mixing in different periodic vortex structures, with two types of
vortices in the unit cell. Fig.~\ref{EMmixture} shows the
difference between the free energy (per vortex line) of some
periodic structures (with up to 9 vortices per unit cell) and the
sum of the energies of two separate perfect hexagonal crystals
formed from SQV and DQV lines. The values in Fig.~\ref{EMmixture}
have been obtained using the planar summation method from
Ref.~\cite{Avilov82}. We see that for all these structures the
energy of mixing is positive -- the lowest energy corresponds to
two simple hexagonal separated lattices. This fact explains the
tendency to form a compact pocket of DQV lines inserted into the
SQV array.

The second feature of the simulation is that the equation of
motion (Eq.~(\ref{eq.of.motion})) does not include any thermal
annealing process. This, in fact, corresponds to the situation in
superfluid $^3$He: Thermal fluctuations are weak, because of the
very low temperatures compared to high energy barriers, associated
with variations in the superfluid velocity field. The above arrays
are therefore not expected to have the optimal structure of the
global free energy minimum.

\subsubsection{Equilibrium arrays}

Experimentally the above simulations correspond to a situation
where the container with superfluid $^3$He-A is slowly accelerated
to rotation at constant temperature below $T_c$. Cooling the
container through $T_c$ in rotation at constant $\Omega$ is
expected to produce a different situation, the global free-energy
minimum. To find this state, we now compare the free energies of
different vortex array configurations. First we note that for
evaluating integrals such as those in Eq.~(\ref{w0.01a}) it is
useful to introduce a vortex-core cutoff at short distances such
that close to the core of the $j$th vortex we replace
\begin{equation}
 \frac{{\bf r-r}_j }{\mid {\bf r-r}_j \mid^2}\;
\rightarrow \; \frac{{\bf r-r}_j}{ \xi_j ^2}~, \label{regul}
 \end{equation}
where $\mid {\bf r-r}_j\mid<\xi_j$. Using Eq.~(\ref{w0.01a}) and
performing integration by part, while neglecting all terms of
order $(\xi_a /r_a)^2$ since $\xi_a\ll r_a$ is assumed, we can
rewrite Eq.~(\ref{w2.01}) in the form
\begin{eqnarray}
\tilde{E}\left({\bf v}_s\right)&=&\sum_{j} \left[ \frac{\rho
_s\kappa _j^2}{4\pi }\left( \ln\left|\frac{r_j^2-R^2}{ R
\xi_j}\right| +\frac{1}{4}\right) \right. \nonumber \\ &&+\left.
\sum_{j\ne l} \frac{\rho _s\kappa _j\kappa _l}{4\pi } \ln\left(
{\mid {\bf r}_l- {\bf r}_j' \mid r_j\over \mid {\bf r}_l- {\bf
r}_j \mid R}\right)\right] \nonumber \\ &&+\sum_{j} \frac{\rho
_s\Omega \kappa _j}{2} (r_j^2-R^2)~. \label{w1.02}
\end{eqnarray}

Next it becomes useful to introduce a vortex lattice energy which
provides a more sensitive measure of the configuration dependent
energy differences. Let us construct the same free energy as in
Eq.~(\ref{w2.01}) but with the superfluid velocity written in the
continuum approximation with ${\bf v}_s={\bf v}_{s0}$, where ${\bf
v}_{s0}$ is defined in Eq.~(\ref{v/contin}). The free energy in
the continuum approximation is then (cf. Eq.~(\ref{energy1/ab}))
\begin{eqnarray}
\tilde{E}\{ {\bf v}_{s0}\} &=&\frac{\rho_s}{ 2}\int _0^R ({\bf
v}_{s0}^2-2 {\bf v}_{s0} {\bf v}_{\rm rot})d^2r \nonumber \\
&=&E_0 \left(\frac{R_{v}}{R}\right)^4\left[ \frac{3}{4}-\ln
\left(\frac{R_{v}}{R}\right) -\left(\frac{R_{v}}{R}\right)^{-2}
\right]~. \label{contin2}
\end{eqnarray}
The difference of the two energies is the Madelung energy,
\begin{equation}
E_M= \tilde{E}\{ {\bf v}_s\}-\tilde{E}\{ {\bf v}_{s0}\}~.
\label{Madelung}
\end{equation}
The Madelung energy represents the vortex lattice contribution and
includes the interactions with the image vortices.
If one neglects the difference ${\bf v}_s - {\bf v}_{s0}$ in the
vortex--free region,
\begin{equation}
E_M = \frac{\rho_s}{2}\int _0^{R_v}({\bf v}_s - {\bf v}_{s0})^2 d^2r~.
\label{Madelung2}
\end{equation}
Within the logarithmic approximation it reduces to the last term in
Eq.~(\ref{energy1/ab}).

An important feature is the interaction with walls. By moving the wall
further from the vortex cluster its influence is reduced and the
vortex cluster starts to approach ``ideal matter'' with more
universal properties, similar to point charges in an ion trap.
Consider the outermost ring of vortex lines in an ideal coaxial
cluster. Nonuniformities in the velocity field around the cluster,
which arise from the discrete nature of the vortex lines in the
outer ring, decay as $\sim \exp(-2\pi \delta r/r_a)$ with distance
$\delta r$ from the outer boundary \cite{Avilov}. Conversely, the
interaction of the outermost vortex ring with its reflection is of
order $\exp(-4\pi d/r_a)$, where $d$ is the distance between the
ring and the container wall. To reduce this effect, we take the
container radius $R$ to be clearly larger than $R_v$ in our
simulation calculations. We then obtain an estimate for the
Madelung energy which does not depend on the interaction of the
particular vortex array with the container wall.

Let us consider an array consisting of only one type of vortex
lines (e.g. $a$). We expect that ideally the minimum energy at
large $N_{a}$ is reached with a configuration with hexagonal
structure in the center of the cluster, circular rings at the
outer boundary, and an intermediate irregular region between these
two regimes. In Eq.~(\ref{Madelung2}) ${\bf v}_{s0}$ is the
average superfluid velocity  within the cluster, equal to the
velocity of rotation (Eq.~(\ref{v/contin})), and thus ${\bf v}_s -
{\bf v}_{s0}$ represents the deviation from the average in the
rotating coordinate system. In an ideal periodic structure this
velocity is also a periodic function of the coordinates and the
integration in Eq.~(\ref{Madelung2}) is reduced to one over a
single unit cell:
\begin{equation}
E_M/N_{a} = \frac{\rho_s}{2}\int_{cell} ({\bf v}_s - {\bf
v}_{s0})^2 d^2r~. \label{Madelung3}
\end{equation}
As a first approximation we may replace the hexagonal lattice cell
with a cylindrical Wigner-Seitz  cell of the same area. The area
per vortex is $b_a^2\sqrt{3}/2=\pi r_a^2$, which gives $ b_a
=1.9046 r_a$ as the relation between the lattice constants $b_a$
of a 2D hexagonal lattice and $r_a$ of a cylindrical WS lattice.
We refer to this approximation as the WS cylinder. The periodicity
of the lattice requires that the component of ${\bf v}_s - {\bf
v}_{s0}$, which is normal to the border of the unit cell, vanishes
at the border. In the rotating coordinate frame the superfluid
velocity has only an azimuthal component and we may write
\begin{equation}
\mid{\bf v}_s - {\bf v}_{s0}\mid= \cases{\kappa_{a} r/( 2\pi
\xi_{a}^2), & if $r<\xi_{a}$\cr
 \kappa_{a} /(2\pi r)-\Omega r ,& if $ \xi_{a}<r<r_{a}$. \cr }
 \label{Madelung4}
 \end{equation}
The Madelung energy can then be written as the energy of a single
vortex
\begin{equation}
E_{M}/N_{a} = {\rho_s \kappa_{a}^2\over 4\pi} \left[\ln{r_{a}\over
\xi_{a}}-\alpha_M\right] \;, \label{WS-cylinder}
\end{equation}
where $\alpha _M$ will be referred to as the Madelung constant.
For the WS-cylinder approximation in Eq.~(\ref{Madelung4}),
$\alpha_M=1/2$. Comparing Eqs.~(\ref{energy1/ab}) and
(\ref{WS-cylinder}) we obtain the expression for the energy density
$\epsilon _{a}$ through the Madelung energy as in
Eq.~(\ref{epsilon}). The Madelung constant $\alpha_M$ and the
Madelung energy can be calculated exactly using the method of
planar summation as in Ref.~\cite{Avilov82}. For an ideal
hexagonal structure it yields $\alpha_M= 0.49877$. This value is
very close to that of the WS-cylinder approximation:
$\alpha_{M}=1/2$.

The Madelung energy in Eq.~(\ref{Madelung}) for large, but finite
$N_{a}$ can be expanded in powers of the small variable $b_a/R_v$:
\begin{equation}
E_{M} = {\rho_s \kappa_{a}^2\over 2\pi} \left[{ 1\over
2}(\ln{r_{a}\over \xi_{a}}-\alpha_M)N_{a}+ 2 \pi \beta
R_v/b_{a}\right] \;. \label{surface1}
\end{equation}
Here the first term within the square brackets is the volume
contribution, and the term proportional to $\beta$ is the surface
energy per unit vortex length
\begin{equation}
\sigma _a =\frac{\rho _s \kappa _a^2\beta}{2\pi b_a} \,.
\label{surfenergy}
\end{equation}
The number of vortex lines in the outermost ring is $2 \pi
R_v/b_{a}$, where the intervortex distance $b_{a}$ is roughly
equal to the lattice constant of the ideal hexagonal structure:
$b_{a}\approx 1.90 r_{a}$. One can check that the factor $\beta$
is independent of the type of vortex as long as the interaction of
the vortex lines with the cylinder wall can be neglected.

To determine the surface energy factor $\beta$, we calculate $E_M$
numerically for different values of $N_{a}$. This requires
minimization of the free energy in Eq.~(\ref{w1.02}) for different
values of $N_{a}$. Here we use the simulated annealing method from
Ref.~\cite{Avilov} where it was applied to calculate the energy of
a large 3D ion cluster in an ion trap. First the vortex lines with
a fictitious mass are heated to some ``temperature'' which is
modeled by a random force. The cooling is modeled by inserting a
small viscous term in a ``Newtonian equation of motion''. The
heating--cooling cycles are repeated, to allow the array to
converge, as a function of time, towards some minimum-energy
configuration with a fixed number of vortex lines.
Eq.~(\ref{w1.02}) is then used to calculate the free energy of the
final converged optimum configuration. Its Madelung energy is
obtained from Eq.~(\ref{Madelung}). Note that plastic deformations
play an important role in making relaxation possible towards the
energy minimum. The (logarithmic) 2D Coulomb interaction can be
regarded as a soft-core potential, in contrast to a hard--sphere
interaction of eg. the Lennard-Jones type. This simplifies the
relaxation of the vortex array towards an optimal structure.
Moreover, in contrast to 3D Coulomb systems, where many crystal
structures (bcc, fcc, hcp) have nearly the same energy, the energy
of a 2D hexagonal crystal is lower than the energy of most other
structures.

\begin{figure}[!t]
  \begin{center}
    \leavevmode
    \includegraphics[width=1.0\linewidth]{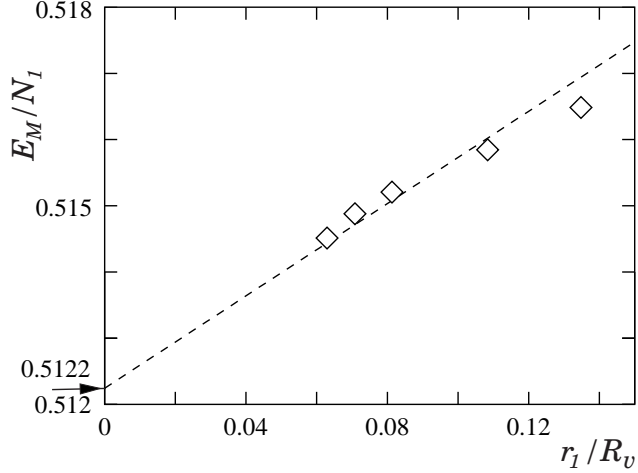}
    \bigskip
    \caption{
Madelung energy $E_M$ for an equilibrium cluster of SQV lines,
plotted as a function of $r_1/R_v = 1/\sqrt{N_1}$ (where $N_1=$
55, 85, 151, 199, and 253). The calculations represent the minimum
value of $E_M/N_1$ (in units of $\rho_s\kappa_0^2/2\pi$) after a
large number of simulated annealing cycles. The container radius
is set at $R=40r_1$ while the core radius is fixed to $\xi
_1/r_1=0.218$. The latter corresponds to SQV lines with $\xi_1
=32\mu$m at $\Omega=0.5$ rad/s. The dashed line represents the
linear dependence of Eq.~(\protect\ref{surface1}) and the fitted
slope gives for $\beta$ the value : $\beta =0.010$. The
extrapolation $E_M/N_1=0.512$ to an infinite vortex array equals
that which one obtains from Eq.~(\protect\ref{WS-cylinder}) for an
ideal hexagonal lattice. }
  \label{RelaxSQV}
  \end{center}
  \end{figure}

\begin{figure}[!t]
  \begin{center}
    \leavevmode
    \includegraphics[width=0.7\linewidth]{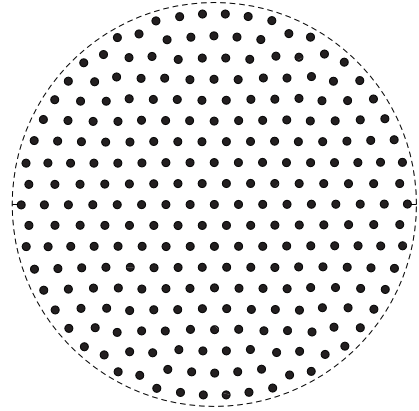}
    \bigskip
    \caption{
Calculated minimum-energy structure of a SQV cluster with $N_1 =
253$ lines. This is one of the examples in
Fig.~\protect\ref{RelaxSQV}. The dashed circle marks the edge of
the cluster at $r=R_v$. The container wall is further out at
$R=2.5 R_v$. This is thus a metastable state with less than the
equilibrium number of vortex lines, but with the cluster itself at
equilibrium density. All calculated clusters in
Fig.~\protect\ref{RelaxSQV} are of this type, to reduce the
influence of the cylinder wall on the simulation calculations. }
  \label{SQV_EquilCluster}
  \end{center}
  \end{figure}

Fig.~\ref{RelaxSQV} shows the Madelung energy $E_M/N_1$  per
vortex as a function of $r_{1}/R_v=1/\sqrt{N_1}$, for arrays
consisting of SQV lines.  We start the simulation from an ideal
hexagonal lattice within a circle $r<R_v$ and after many
$(\sim10^2)$ heating--cooling cycles we arrive at the structure
which corresponds to a stable minimum of Eq.~(\ref{w1.02}). By
repeating this procedure for different values of $R_v$ (or $N_1$),
we obtain $\beta$ in Eq.~(\ref{surface1}) from the fitted slope in
Fig.~\ref{RelaxSQV}: $\beta =0.010$. The calculated optimal
structure of these vortex clusters (Fig.~\ref{SQV_EquilCluster})
approaches that of an ideal hexagonal lattice in the center, but
then distorts to a quasi-circular structure towards the outer
boundary, with a transition region between these two extremes.

\begin{figure}[!t]
  \begin{center}
    \leavevmode
    \includegraphics[width=1\linewidth]{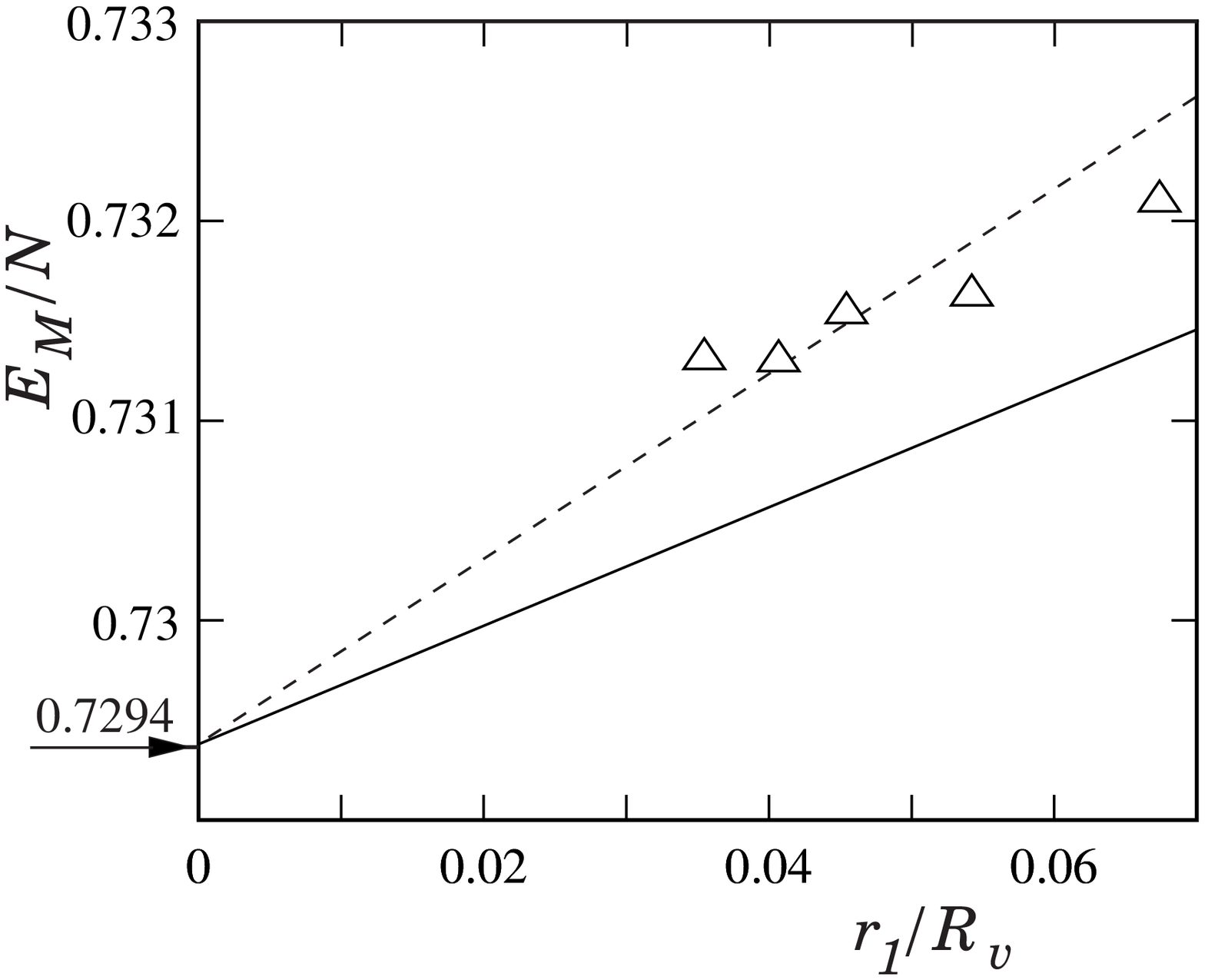}
    \bigskip
    \caption{
Calculated Madelung energy per circulation quantum, $E_M/N$, for
segregated coaxial minimum-energy clusters, composed of SQV and
DQV lines, plotted as a function of the inverse of the normalized
cluster radius: $r_1/R_v = 1/\sqrt{N}$. In the minimum energy
configuration the DQV lines are in the center. Here the total
circulation of the DQV lines has been taken to be equal to that of
the SQV lines (i.e. $N_1= 2N_2 = N/2$). $E_M$ is expressed in
units of $\rho_s\kappa_0^2/2\pi$ and is calculated for five
cluster radii $N_2=55,\, 85,\, 121,\, 151,$ and 199. The
parameters are $\xi _1/r_1=0.218$, $\xi _2/r_1=0.236$, which
correspond to $\xi _1=32\;\mu$m and $\xi_2=49\;\mu$m at $\Omega =
0.5$ rad/s. The container radius is set to $R=40 \, r_1 = 5.8$ mm.
The arrow indicates the sum of the Madelung energies in
Eq.~(\protect\ref{surface1}) for infinite SQV and DQV  lattices,
$(E_{M1}+E_{M2})/N$ in the limit $N\to \infty$, when the surface
energy can be neglected. The solid line has the same slope as the
line in Fig.~\protect\ref{RelaxSQV} and represents the surface
energy of the outer boundary of a cluster with SQV lines. The
dashed line is a linear fit through the data points. }
  \label{SimulPlot2}
  \end{center}
  \end{figure}

\begin{figure}[!t]
  \begin{center}
    \leavevmode
    \includegraphics[width=0.7\linewidth]{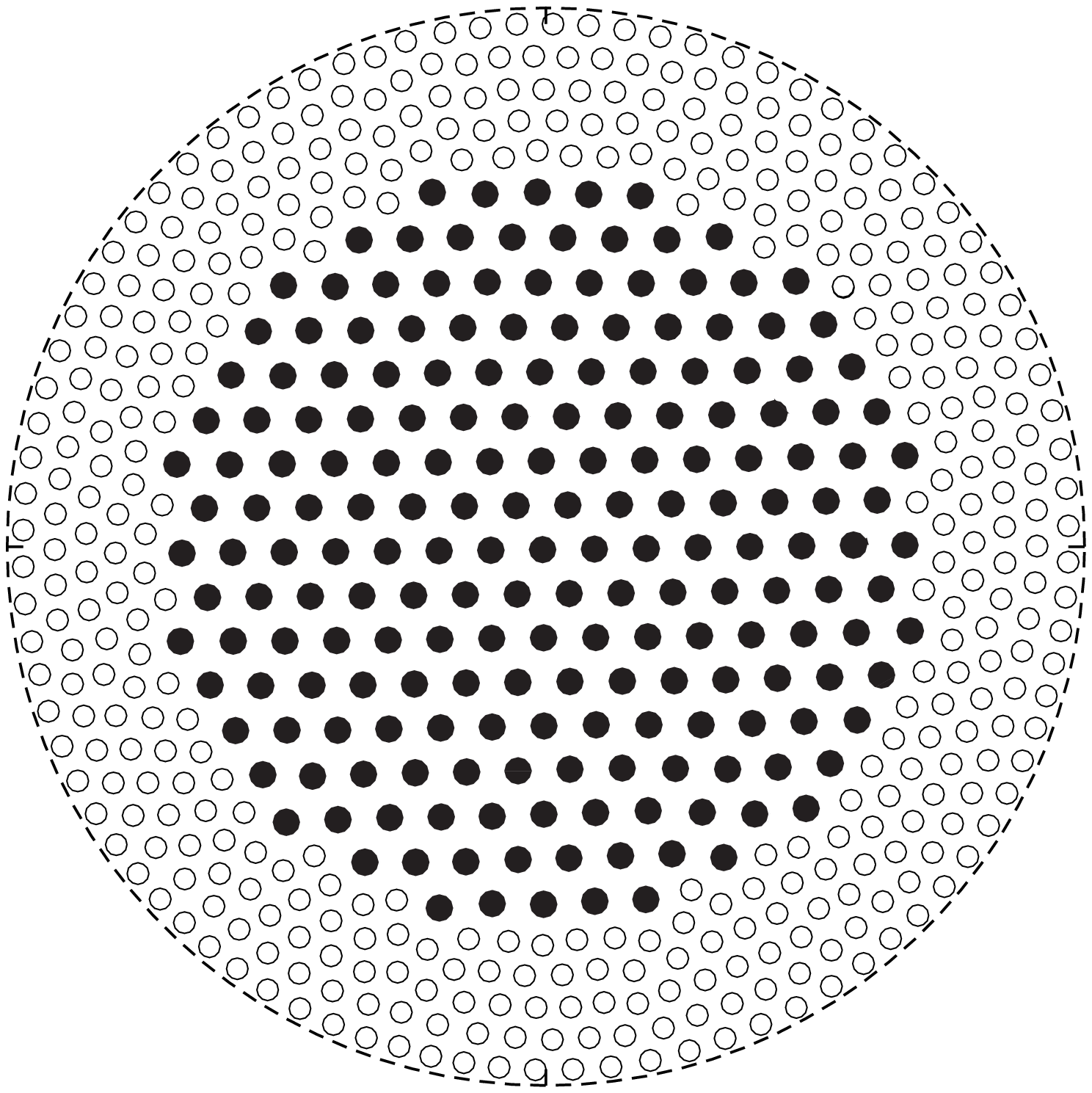}
    \bigskip
    \caption{
Calculated minimum-energy structure of a cluster with $ N_2 = 199$
and $N_1 = 2 N_2$ lines. This is one of the examples from
Fig.~\protect\ref{SimulPlot2}. The dashed circle denotes the edge
of the cluster at $r=R_v$ while the container wall is at $R=1.4
R_v$. }
  \label{CoExCluster}
  \end{center}
  \end{figure}

Similar simulations for the equilibrium state of a cluster with
two coexisting vortex species are more delicate. First, we have
already noted that the energy of a mixture of SQV and DQV lines is
higher than that of two segregated domains. Secondly, it follows
from Eq.~(\ref{surfenergy}) that the surface energy of a DQV
cluster is $2 \sqrt{2}$ times larger than that of a SQV cluster
with the same value of $R_v$, i.e., $\sigma _2/\sigma _1
=2\sqrt{2}$ because $\kappa_2=2\kappa_1$ and $r_2=\sqrt{2} \,
r_1$. Hence, the only configurations, which were found to converge
towards a stable minimum-energy state, are coaxial clusters with
the DQV lines in the center, surrounded by an outer annulus of SQV
lines. In Fig.~\ref{SimulPlot2} the Madelung energy $E_M/N$ per
circulation quantum of such clusters has been  plotted vs.
$r_{1}/R_v=1/\sqrt {N}$, where $N=N_1+2N_2$ is the total number of
circulation quanta. We take here $N_1= 2N_2$ and thus the
interface separating the two types of vortex lines lies at
$r\approx R_v/\sqrt{2}$. The optimal structure
(Fig.~\ref{CoExCluster}) turns out to consist of a nearly ideal
hexagonal lattice of DQV lines, which extends all the way from the
center to the inner interface. At the inner interface the mismatch
in periodicity between the two vortex species, $b_2=\sqrt{2} b_1$,
creates an abrupt break, but the smaller SQV lines very
effectively compensate for the faults on the boundary of the
central hexagonal DQV crystal. Also the SQV lines fulfill the
boundary condition and on approaching the outer boundary of the
cluster a perfect circular ring structure is established.

Simulation attempts on the inverse initial state, namely on
segregated coaxial domains but with the SQV lines in the center,
were not successful. Such calculations are slow in convergence
since they start to drift towards the ideal structure where the
SQV lines are in the outer periphery. The same applies to
asymmetric starting configurations, such as the one where the
cluster is divided along its diameter into equal halves with only
SQV and DQV lines in each. This arrangement does not minimize the
outer surface energy or allow for the presence of a symmetric
hexagonal crystal in the center of the array.

We can use the results in Fig.~\ref{SimulPlot2} to estimate the
surface energy of the inner interface $\sigma_{12}$. The dashed
line shows a fit to the calculations: $E_M(DQV+SQV)/N = 0.7294 +
0.0462 \, r_1/R_v$ (in units of $\rho_s \kappa^2/2\pi$). The same
fit to the SQV clusters in Fig.~\ref{RelaxSQV} gives
$E_M(SQV)/N=0.5122+0.0348 \, r_1/R_v$. Since the outer shell in
both cases consists of SQV lines only, we can associate the
difference in slopes with the surface energy of the inner
interface between the SQV and DQV domains. This gives $\delta
E_{12}/N=0.012 \, r_1/R_v$. Taking into account the length $2\pi
R_2 \approx 2\pi R_v/\sqrt{2}$ of the interface, we obtain
$\sigma_{12}/\sigma_1\approx 0.48$. Remembering that
$\sigma_{2}/\sigma_1=2.83$, we find that
$\sigma_2>\sigma_1+\sigma_{1,2}$. Thus the contact angle between
the SQV domain and the vortex-free CF annulus has to be zero: the
SQV domain ``wets'' completely the vortex-free region. The most
preferable configuration is therefore one where the DQV domain is
completely surrounded by SQV lines.

\begin{figure}[t]
  \begin{center}
    \leavevmode
    \includegraphics[width=0.8\linewidth]{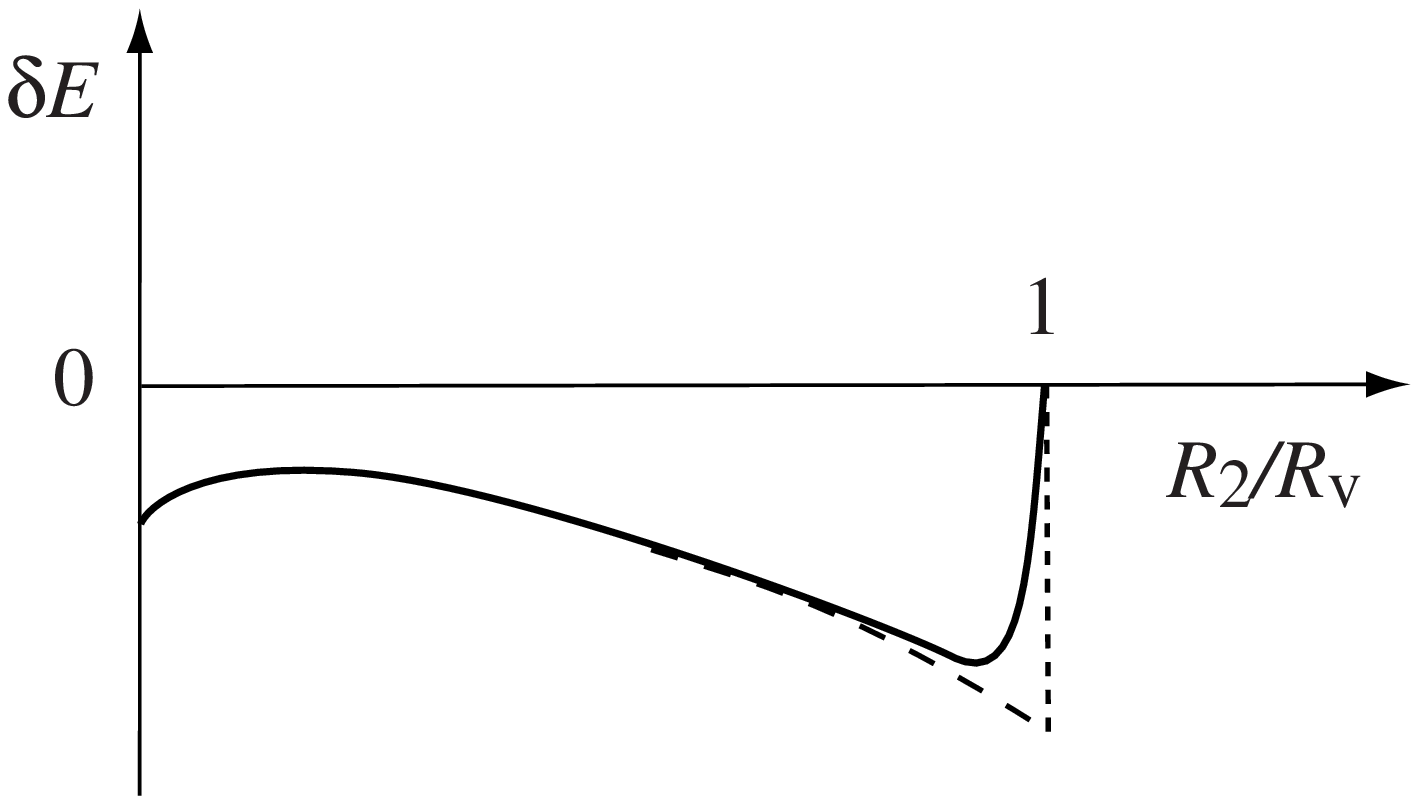}
    \bigskip
    \caption{
Free energy difference $\delta E$ between a DQV cluster with
radius $R_2$ and one which has an outer surface layer of SQV lines
in the region $R_2<r<R_v$, when $\Omega >\Omega ^*$.  Within the
idealized model of Eq.~(\protect\ref{surface1}), there exists a
step-like energy reduction by $2\pi R_v(\sigma _2-\sigma _1-\sigma
_{12})$ {\it (dotted line)}, when a thin layer of SQV lines is
placed on the outer surface of a DQV cluster. Decreasing the
radius $R_2$ of the DQV domain further results in an increase in
the free energy {\it (dashed curve)} until it reaches a maximum
followed by another minimum at $R_2=0$. The {\it solid curve}
indicates schematically the dependence with a SQV surface layer of
finite width. } \label{coexist}
\end{center}
\end{figure}

Interestingly the reduced surface energy with SQV lines makes the
coexistence of DQV and SQV lines possible even in a region of the
SQV -- DQV phase diagram where the global energy minimum
corresponds to that of DQV lines only. The energy of a cluster
consisting of DQV lines only is given by Eq.~(\ref{surface1}),
\[
E_M^{(2)} =\frac{\rho _s\kappa _2^2R_v^2}{4\pi r_2^2}\left(\ln
\frac{r_2} {\xi _2} -\alpha _M\right) +2\pi R_v\sigma_2 \;,
\]
while the energy of a cluster with a central DQV domain surrounded
by SQV lines is
\begin{eqnarray*}
E_M^{(21)} &=&\frac{\rho _s\kappa _2^2R_2^2}{4\pi r_2^2}\left(\ln
\frac{r_2} {\xi _2} -\alpha _M\right)\\ &&+ \frac{\rho _s\kappa
_1^2(R_v^2-R_2^2)}{4\pi r_1^2}\left(\ln \frac{r_1} {\xi _1}
-\alpha _M\right)\\ && +2\pi R_v\sigma _1 +2\pi R_2\sigma _{12}\;.
\end{eqnarray*}
$E_M^{(21)}$ as a function of $R_2$ has a minimum at $R_2=0$ for
$\Omega <\Omega ^*$, when a SQV cluster is more favorable.
However, due to a finite $\sigma _{12}$, there is still a minimum
at $R_2=0$ even when $\Omega >\Omega ^*$, i.e. when a DQV cluster
is more favorable. In this second case another minimum appears at
$R_2=R_v$ which corresponds to a DQV cluster surrounded by a thin
surface layer of SQV lines. Indeed, the energy $E_M^{(21)}$ is a
decreasing function of $R_2$ in the region
 $R_2\sim R_v$. For $R_2=R_v$ it is smaller than $E_M^{(2)}$ because
$\sigma _2>\sigma _1+\sigma _{12}$. The difference $\delta
E=E_M^{(21)}-E_M^{(2)}$ for $\Omega >\Omega ^*$ is shown in
Fig.~\ref{coexist} as a  function of $R_2/R_v$. The two local
minima at $R_2=0$ and $R_2\approx R_v$ have the same energy when
\begin{equation}
\Omega^{\dag}  =\Omega ^*\left[ 1+\frac{8\pi \beta
r_1^2}{R_vb_1}\frac{\sigma _{12}} {\sigma _1}\right]\; .
\label{newcondition}
\end{equation}
Here $\Omega ^*$ is determined by Eq.~(\ref{Omega*}) where,
however, $e$ should be replaced with the more accurate expression
$e^{2\alpha _M}$ of an ideal vortex lattice. $\Omega^{\dag}$ is
only slightly higher than $\Omega ^*$ because of the large value
of $R_v/r_1$.

Eq.~(\ref{newcondition}) states the condition for the global
equilibrium between SQV and DQV lines. However, because of the
minimum at $R\approx R_v$, the model allows for the coexistence of
the two vortex species: A ``monolayer'' in the form of an outer
circle of SQV lines could surround a DQV cluster, when $\Omega
>\Omega ^*$. Evidently, our idealized model only applies if the
vortex layer is at least a few inter-vortex distances wide. With
this warning we note that the qualitative behavior of $\delta E$
is shown in Fig.~\ref{coexist} by the solid line, when an outer
boundary layer of SQV lines is present.

Summarizing our analysis we conclude that the minimum energy
configuration for coexisting SQV and DQV lines in a rotating
cylinder can only be established by considering discrete vortex
arrays. The optimal configuration (1) consists of segregated
domains, (2) has cylindrical symmetry (3) with the SQV lines
forming the outer boundary of the cluster.

\section{Experimental method}

The structure of the vortex cluster in the rotating container is
probed with cw NMR. It allows one to determine the average radial
composition, ie. $N_1(r^2)$ and $N_2(r^2)$. The NMR method is
based on a measurement of the absorption spectrum to which SQV and
DQV lines each contribute a characteristic satellite peak. These
satellites have frequency shifts which are different from each
other and from the bulk liquid NMR line \cite{Ruutu3}. Both the
peak height and the integrated intensity of the satellites are
proportional to the total number $N_a$ of vortex lines of the
species $(a)$ and can be calibrated to give the absolute value of
$N_a$.

To measure the composition of a vortex cluster we monitor the peak
height of a satellite as a function of the rotation velocity
$\Omega$ during slow deceleration, when the cluster expands and
the outermost vortex lines annihilate one by one at the cylinder
wall \cite{Ruutu2}. The peak height as a function of $\Omega$
gives the number of the annihilating vortex lines as a function of
$r^2$. We can thus reconstruct the cumulative number of both types
of vortices $N_a(r^2)$, which gives the number of vortex lines of
type $(a)$ inside a radius $r$. Indeed, for reconstructing the
radial distributions only one of the peak height dependences
$N_a(r^2)$ needs to be measured. The requirement of solid-body
rotation of the superfluid component within the vortex cluster
gives the relation: $N_1(r) + 2N_2(r) = 2\pi r^2 \Omega
/\kappa_0$.

The measurements have been performed with two cylindrical
containers, one fabricated from epoxy and the other from fused
quartz \cite{Ruutu4}. Both are right circular cylinders 7 mm long
and  $R = 2.5$ mm in radius. They are closed off, except for an
orifice of 0.5 mm in diameter, located in the center of the flat
bottom plate. The orifice connects to a long tubular channel which
provides via its liquid $^3$He column the thermal contact to the
refrigerator. The $^3$He-A  temperature in the container is
determined from the calibrated temperature dependent frequency
shift of the bulk liquid NMR absorption peak.

Two controlled ways exist for preparing a vortex cluster: 1) The
sequential addition of vortex lines one by one to the cluster by
accelerating $\Omega$ slowly (in the presence of a constant
critical velocity), and 2) by cooling in rotation through $T_c$.
One might expect in the latter case that a sufficiently slow cool
down through $T_c$, one which approaches the adiabatic limit, will
lead to an equilibrium state, where the number and type of vortex
lines as well as the configuration of the cluster adjust such that
the total energy becomes minimized. This has been experimentally
checked with respect to the type of lines \cite{Parts1} and the
number of lines \cite{Ruutu2}.

In $^3$He-A the transition between the SQV and DQV lines occurs in
the presence of an applied magnetic field \cite{SoftCore} as a
function of $\Omega$ \cite{Parts1}. Experimentally we find that
there exists a region of coexistence around this transition, ie.
the transition has a width as a function of $\Omega$  within which
both species of vortex lines are present simultaneously, but where
their relative amounts depend on $\Omega$. Ideally the transition
should be sharp and therefore our experimental situation cannot
represent the homogeneous adiabatic limit. Nevertheless, in the
coexistence regime the energy difference between the two vortex
species is small, and, in principle, it can be smaller than the
energies associated with the vortex cluster configuration.
Therefore, in addition to trying to identify what type of
inhomogeneity controls the width of the transition, we may
investigate the vortex-array configurations in the coexistence
regime and ask whether these energy contributions are minimized in
the transition.

The sequentially formed coexistence clusters are more obvious.
Only one type of cluster is possible: Due to their much higher
critical velocity SQV lines can only be formed by cooling through
$T_c$ in rotation. Thus by cooling slowly through $T_c$ at low
$\Omega$ an equilibrium cluster of SQV lines is formed, to which
DQV lines can be added sequentially by increasing $\Omega$ slowly
at constant temperature below $T_c$.

The first observation from our measurements was that the radial
distributions of the two vortex species turned out to be different
in the equilibrium clusters and the sequentially constructed
states. This means that in superfluid $^3$He a common cluster
configuration is not realized, but metastable states are long
lived and the structure of a vortex cluster depends on how it has
been prepared.

\section{Measurements}\label{Chap:Exp}
\subsection{Coexistence regimes of SQV and DQV}
\subsubsection{Dependence on rotation while cooling through $T_c$}
\label{Sec:RotDependence}

\begin{figure}[bt!!!]
\begin{center}
\leavevmode
\includegraphics[width=0.9\linewidth]{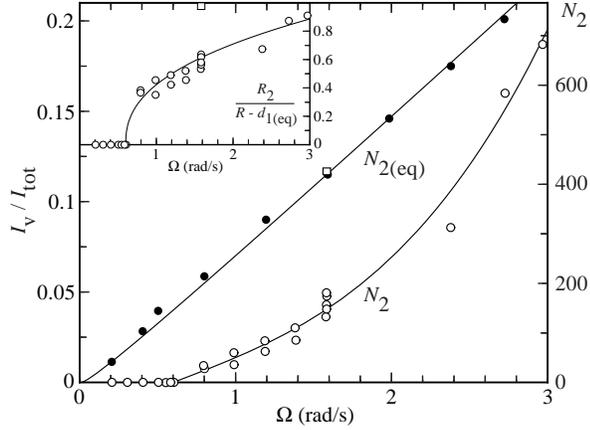}
\bigskip
\vfill \caption{Coexistence of SQV and DQV lines in the high-field
(dipole-unlocked \protect\cite{SoftCore}) regime as a function of
rotation velocity $\Omega$. $(\circ)$: Normalized DQV satellite
intensity $I_v/I_{tot}$ plotted as a function of the rotation
velocity at which the sample was cooled through $T_c$. On the
right vertical axis the intensity has been expressed in terms of
line numbers, $N_2(\Omega)$, by means of the calibration,
$N_{2(eq)}$.   At low $\Omega$ only SQV lines are formed and
$N_2(\Omega < 0.6$ rad/s) = 0. Above 0.6 rad/s the coexistence
regime starts, where the fraction of the total circulation in DQV
lines increases with $\Omega$, until well above 3 rad/s no more
SQV lines are formed. As a guide for the eye, the solid curve
represents $I_v/I_{tot} = 0.033(\Omega - 0.6) + 0.0084 (\Omega -
0.6)^3$. The cool-down rate at $T_c$ has been kept at
$|dT/dt|_{T_c} \lesssim 5\; \mu$K/min, except for the data point
marked with an open square $(\Box)$, for which the cooling rate
was $(dT/dt)_{T_c} = -18 \; \mu$K/min. $(\bullet)$: Calibration
measurement of $I_v/I_{tot}$ with the equilibrium number of DQV
lines, $N_{2(eq)}$. The fitted result gives $I_v(\Omega)/I_{tot} =
2.74 \cdot 10^{-4} \; N_{2(eq)}(\Omega)\approx 0.0806 \; \Omega \;
(1-0.1/ \sqrt{\Omega})$ ($\Omega$ in rad/s).~The~{\it inset}~shows
the radius of the central DQV cluster in relative units in the
coexistence regime, together with a guide for the eye:
$R_{2}/(R-d_{1(eq)}) = 0.618 \; (\Omega - 0.6)^{0.41}$. Other
conditions: $ P = 29.3$ bar, $H = 9.91$ mT, the satellite
intensities have been measured at $T= 0.81\;T_c$.}
\label{EquilMixArray}
\end{center}
\end{figure}

When the rotating sample is cooled in the NMR field through $T_c$,
an equilibrium vortex state with only SQV lines is formed at low
rotation, $\Omega \leq 0.6 $ \cite{Parts1}. Above this velocity
also DQV lines are created. Their relative number increases with
$\Omega$ until above 3 rad/s no more SQV lines are observed. The
number of DQV lines is shown in the coexistence regime as a
function of $\Omega$ in Fig.~\ref{EquilMixArray}, where each data
point $(\circ)$ represents $I_v/I_{tot}$, the integrated NMR
absorption $I_v$ of the DQV satellite normalized to the total
absorption $I_{tot}$ in the NMR spectrum.

The normalized absorption $I_v(\Omega)/I_{tot}$ can be translated
to vortex line numbers $N_2(\Omega)$  by comparing to an
equilibrium state with only DQV lines. The equilibrium number of
vortex lines (of type $a$ only) is given by
\begin{equation}
N_{a(eq)}(\Omega) =\pi (R-d_{a(eq)})^2 \; {2\Omega \over \kappa_a}
\; \propto \; {{I_v (\Omega)} \over I_{tot}}\;\;, \label{e:N_2eq}
\end{equation}
where the width $d_{a(eq)}$ of the equilibrium CF annulus is
defined in Eq. (\ref{e:deq}). The calibration measurement,
$N_{2(eq)}(\Omega)$, is plotted $(\bullet)$ in
Fig.~\ref{EquilMixArray}. In practice, the equilibrium number of
vortex lines is equal to that at the annihilation threshold
\cite{Ruutu2}. In Fig.~\ref{EquilMixArray} the data for
$N_{2(eq)}(\Omega)$ were measured by first slowly increasing
$\Omega$ from zero to some high velocity $\Omega > 3.5$ rad/s at
$T=0.81\; T_c$ and by then decelerating sequentially to lower
$\Omega$ values and recording there at constant $\Omega$ the
respective value of $I_v(\Omega)/I_{tot}$.  The difference between
the two curves in Fig.~\ref{EquilMixArray} gives the number of SQV
lines in the metastable coexistence regime: $N_1(\Omega) =
2\;[N_{2(eq)}(\Omega) \; - \; N_2(\Omega)]$.

\begin{figure}[bt!!!]
\begin{center}
\leavevmode
\includegraphics[width=0.9\linewidth]{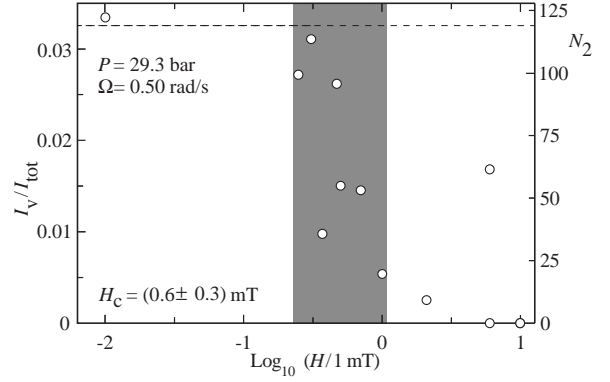}
\bigskip
\vfill \caption{Coexistence of SQV and DQV lines as a function of
magnetic field $H$ below  0.6 rad/s: Normalized DQV satellite
intensity $I_v/I_{tot}$ plotted vs the field $H$, which was
applied during cooling through $T_c$ at  0.5 rad/s. On the right
vertical axis the NMR absorption intensity has been converted to
DQV vortex numbers, $N_2(H)$, using the calibration in
Fig.~\protect\ref{EquilMixArray}. The horizontal dashed line
represents the equilibrium number $N_{2(eq)}$ of DQV lines. The
shaded region marks the width of the coexistence regime around the
critical field, $H_c \approx 0.6$ mT. In this measurement the
field was increased  to the NMR value 9.91 mT in the superfluid
state below $T_c$, while the sample was cooling to $0.81 \; T_c$,
where the intensity $I_v/I_{tot}$ was measured. Other conditions
were the same as in Fig.~\protect\ref{EquilMixArray}. In the
transition through $T_c$ the equilibrium value of total
circulation is created and thus at high field $H
> H_c$, where $N_2 \rightarrow 0$ in accordance with the results in
Fig.~\protect\ref{EquilMixArray}, the deficit from $N_{2(eq)}$
represents the SQV lines. }
\label{FieldDepend0.5rad/s}
\end{center}
\end{figure}

The critical rotation velocity $\Omega ^*$ for the transition from
SQV to DQV lines lies thus according to Fig.~\ref{EquilMixArray}
within the interval 0.6 -- 3 rad/s. An estimate for $\Omega ^*$
can be worked out from Eq.~(\ref{Omega*}) using the normalized
satellite intensity $I_v / I_{tot}$. It measures the effective
vortex-core area compared to that of the WS cell \cite{SoftCore},
or
\begin{equation}
{I_v \over I_{tot}} \approx \left( \xi_a \over r_a \right)^2 =
{{2\pi \xi_a ^2} \over \kappa_a} \; \Omega \;.\label{I_v}
\end{equation}
For the sake of comparison we neglect the temperature dependence
of the core radii \cite{SoftCore} and use the best values for the
satellite intensities, which come from measurements at $0.50\;T_c$
(11.7 mT, and 33.9 bar): In the equilibrium state the coefficient
in front of $\Omega$ in Eq.~(\ref{I_v}) was found to be 0.094 for
SQV and 0.114 for DQV lines. These values give for the soft core
radii $\xi_1 \approx 32 \; \mu$m and $\xi_2 \approx 49 \; \mu$m.
From Eq.~(\ref{Omega*}) we now get $\Omega^* \approx 2.8$ rad/s,
in qualitative agreement with the experiment. The estimate for
$\Omega ^*$ in Eq.~(\ref{Omega*}) is based on the differences in
quantization number, core size, and the crude approximation for
vortex energy in Eq.~(\ref{epsilon}). A more sophisticated
numerical calculation in Ref.~\cite{Thuneberg} places $\Omega^*$
in the interval 0.54 -- 1.4 rad/s.

\subsubsection{Dependence on magnetic field while cooling through $T_c$}

In addition to $\Omega$, the second important variable of the
vortex phase diagram is the applied magnetic field $H$
\cite{Pekola}: When it exceeds the equivalent of the spin-orbit
coupling, as is the case in our NMR measurements, the SQV and DQV
structures display  well-separated vortex cores, with $r_a \gg \xi
_a$ \cite{SoftCore}. At low fields the spin-orbit interaction
dominates and these vortex structures are not stable. Instead here
vorticity appears in the form of extended periodic vortex textures
with no core or a much expanded core, such that $r_a \sim \xi_a$.
The dominant low-field vortex texture has a square lattice and a
unit cell with four quanta of circulation \cite{Parts1,Thuneberg}.
These vortex textures are separated from the high-field structures
by a transition of first order.

The transition between the low and high field regimes as a
function of $H$ is illustrated by the measurements in
Fig.~\ref{FieldDepend0.5rad/s}. Our NMR is performed at a fixed
frequency in the high field limit. Therefore the measurements in
Fig.~\ref{FieldDepend0.5rad/s} have to follow a particular
routine. First, the container is cooled through $T_c$ at 0.5
rad/s, with the field $H$ adjusted to the value given on the
horizontal axis. Next, below $T_c$ the field is increased to the
NMR value of 9.91 mT. During the field sweep the low field vortex
texture undergoes a topological transition to the high-field
(dipole-unlocked) DQV structure, with well-separated vortex cores.
This transition is not revealed in our measurements, if the field
sweep is conducted at high temperatures ($T > 0.8\;T_c$). However,
the SQV has different topology \cite{SoftCore} and any vortices of
this kind will remain singly quantized during the field sweep.
After the field sweep, the container is cooled to lower
temperatures, to improve the resolution in the NMR measurement of
the vortex satellites, and the DQV satellite intensity is
measured.

As shown in Fig.~\ref{FieldDepend0.5rad/s}, we find a critical
field $H_c \approx 0.6$ mT. Above $H_c$ the SQV lines dominate, as
required when $\Omega < 0.6$ rad/s. Below $H_c$ only DQV lines are
found, which represent the transformed low-field vortex textures
after the field sweep. Centered around $H_c$ there is a
coexistence regime with a width of 0.5 mT, where both types of
vortex lines are present. Within the coexistence regime, the
relative abundance of SQV lines increases continuously with
increasing $H$, similar to the situation in
Fig.~\ref{EquilMixArray} as a function of $\Omega$.
\begin{figure}[bt]
\begin{center}
\leavevmode
\includegraphics[width=1\linewidth]{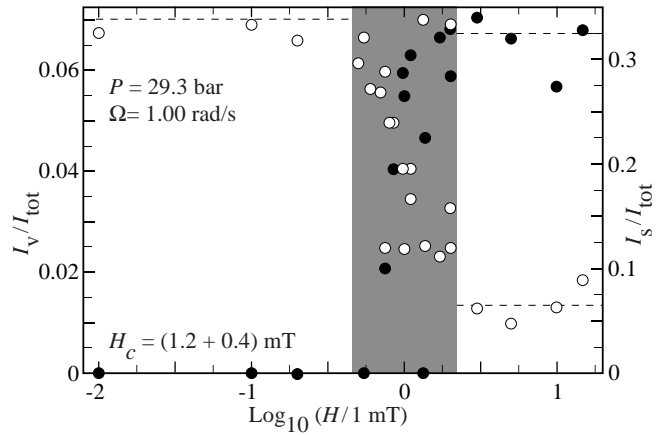}
\bigskip
\vfill \caption{Coexistence of SQV and DQV lines as a function of
magnetic field above  0.6 rad/s: The normalized intensities
$I_v/I_{tot}$ of the DQV satellite $(\circ$; left vertical axis)
and $I_s/I_{tot}$ of the SQV satellite $(\bullet$; right vertical
axis) have been plotted as a function of the magnetic field $H$,
which was applied during cooling through $T_c$ at 1 rad/s. The
horizontal dashed line on the left  represents the equilibrium
value expected for DQV lines, after a superfluid transition at low
fields ($H < H_c$) and a subsequent increase of the field to 9.91
mT. On the right ($H > H_c$), the dashed lines give the mean
values for the SQV and DQV lines: $N_1 \approx 440$ and $N_2
\approx 48$. The shaded region marks the width of the transition
regime around the critical field, $H_c \approx 1.2$ mT. The
coexistence regime includes both the shaded and the high field
regions, $H > H_c$. The same conditions apply as in
Fig.~\protect\ref{EquilMixArray}, including the calibration for
the DQV satellite intensity. The calibration for the SQV satellite
intensity is $N_{1(eq)}(\Omega) \propto I_s(\Omega)/I_{tot}
\approx 0.43 \; \Omega \; (1-0.089/ \sqrt{\Omega})$ ($\Omega$ in
rad/s). } \label{FieldDepend1rad/s}
\end{center}
\end{figure}

The same behavior is verified in Fig.~\ref{FieldDepend1rad/s} at a
higher rotation velocity of 1 rad/s. On comparing
Figs.~\ref{FieldDepend0.5rad/s} and \ref{FieldDepend1rad/s} it is
noted that both $H_c$ and the width of the coexistence regime
increase with $\Omega$ \cite{Parts1}. In the measurements of
Fig.~\ref{FieldDepend1rad/s} both the DQV and SQV satellite
intensities were determined separately. On the low field side only
DQV lines are found. On the high field side both types of lines
are formed, with a 4 times larger share of the circulation carried
by the SQV lines. This is the same ratio as was already seen in
Fig.~\ref{EquilMixArray} at 1 rad/s.

In a magnetic field the superfluid transition splits into $T_{c1}$
and $T_{c2}$. The former is the transition from the normal to the
A$_1$ phase and the latter from A$_1$ to the A$_2$ phase, which is
the regular $^3$He-A. The thermal width of the A$_1$ region is
narrow, approximately 0.6 $\mu$K at 10 mT. We might wonder what
effect the intermediate A$_1$ phase has on the DQV to SQV ratio
above $H_c$ in Figs.~\ref{EquilMixArray} --
\ref{FieldDepend1rad/s}: Are more changes to be expected at fields
much larger than those shown in Figs.~\ref{FieldDepend0.5rad/s}
and \ref{FieldDepend1rad/s}? We find that at least a cool down in
60 mT field at 1.4 rad/s gives the same result as at 10 mT.

\subsubsection{Dependence on cooling rate at $T_c$}

Surprisingly in Fig.~\ref{EquilMixArray} one data point deviates
from the typical behavior (denoted with $(\Box)$ at 1.6 rad/s).
The only distinction that can be appended to its measurement,
compared to other data points at the same $\Omega$, is a larger
cooling rate during the superfluid transition: The five data
points measured at cooling rates $|dT/dt|_{T_c} =$ 1 --- 5
$\mu$K/min give for the relative amount of DQV lines $N_2
/N_{2(eq)} \approx 0.38$ while the anomalous point at a cooling
rate of 18.5 $\mu$K/min falls on the equilibrium curve
$N_{2(eq)}$, with no SQV lines detected at all.

\begin{figure}[bt!!!]
\begin{center}
\leavevmode
\includegraphics[width=0.9\linewidth]{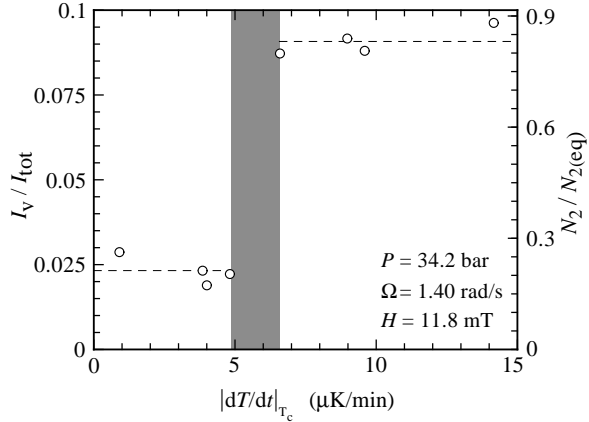}
\bigskip
\vfill \caption{Influence of cooling rate at $T_c$ on the relative
number of SQV and DQV lines. The left vertical axis gives the
normalized DQV satellite intensity $I_v/I_{tot}$ as a function of
cooling rate during the transition through $T_c$ at 1.4 rad/s. The
satellite intensities are measured at (0.49 --- 0.51) $T_c$, where
the calibration for the DQV line satellite is $N_{2(eq)}(\Omega)
\propto I_v (\Omega) /I_{tot} \approx 0.078 \; \Omega \; (1- 0.11/
\sqrt{\Omega})$ ($\Omega$ in rad/s). In this expression $I_v$
includes only the primary satellite peak (the secondary peak,
which is also visible at these low temperatures, is not included
in the absorption). On the right vertical axis the calibration has
been used to list the fraction of the total circulation in DQV
lines, $N_2/N_{2(eq)}$, while the rest, $1-N_2/N_{2(eq)}$, is
carried by SQV lines. } \label{CoolRateDepend}
\end{center}
\end{figure}

Additional evidence is shown in  Fig.~\ref{CoolRateDepend} on how
a faster cooling rate at $T_c$  shifts vorticity from SQV to DQV
lines. In these measurements a fused quartz container was used in
which $^3$He-A could be supercooled to below $0.49 \; T_c$. This
feature improves the measuring resolution since at lower
temperatures the different vortex satellites are separated by
larger shifts from each other in the absorption spectrum
\cite{Ruutu3}. When the cooling rate in Fig.~\ref{CoolRateDepend}
at 1.4 rad/s reaches 5 $\mu$K/min a change over takes place from
SQV to DQV lines: at lower rates only one fourth of the total
circulation is in DQV lines (which is the same fraction as in
Fig.~\ref{EquilMixArray}), while above 6 $\mu$K/min, only few  SQV
lines  are formed. Similar measurements at 0.6 rad/s with cooling
rates up to 16 $\mu$K/min did not yet produce a DQV satellite of
measurable height.

The quartz container in the measurements of
Fig.~\ref{CoolRateDepend} was thermally connected to the
refrigerator via an orifice of 0.5 mm in diameter. Later
measurements with another quartz cylinder with a small and long
channel of 0.4 mm diameter displayed appreciably smaller fractions
of SQV lines, also in the regime $\Omega < 0.6$ rad/s where $N_2 =
0$ in Fig.~\ref{EquilMixArray}. The narrow and long channel
increases the thermal resistance between the sample container and
the refrigerator, and leads to rapid irreversible
normal-to-superfluid transitions, where the transition velocity
cannot reliably be controlled externally. This suggests that a
large cooling rate and nonequilibrium conditions enhance the
formation of DQV lines at the expense of SQV lines.

There is one data point also in Fig.~\ref{FieldDepend0.5rad/s} at
6 mT with a larger fraction of DQV lines than what one would
expect on the basis of the general behavior. In the measurements
as a function field (Figs.~\ref{FieldDepend0.5rad/s} and
\ref{FieldDepend1rad/s})  the cooling rate at $T_c$ cannot be
monitored accurately, since the NMR measurement is inoperative at
that stage. Therefore a cooling routine was worked out which
should have yielded rates of less than 5 $\mu$K/min. However,
without the NMR record it is not possible to state whether a
mistake  occurred in the cooling routine of the anomalous data
point in Fig.~\ref{FieldDepend0.5rad/s}.
\begin{figure}[!t!!!]
  \begin{center}
    \leavevmode
    \includegraphics[width=0.9\linewidth]{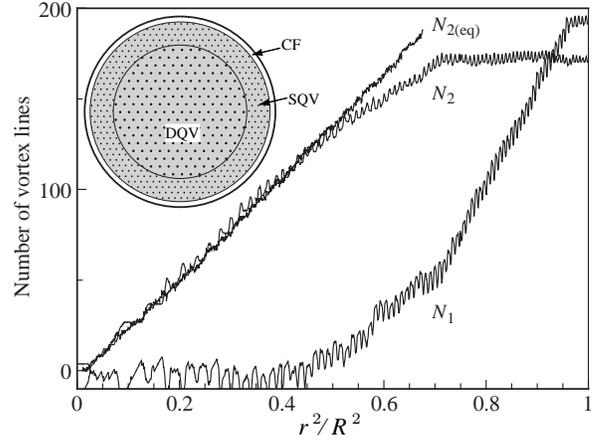}
    \bigskip
  \caption{
Radial distributions $N_1(r^2)$ and $N_2(r^2)$  after cooling
through $T_c$ at 1 rad/s and 1.1 mT. The deceleration from 1 rad/s
to zero is performed at $0.81\;T_c$ and 9.91 mT. Other conditions
are the same as in Fig.~\protect\ref{Glass}. $N_{2(eq)}$ shows for
comparison the deceleration for an array consisting of only DQV
lines. It is prepared by cooling through $T_c$ at $\Omega > 1.5$
rad/s in zero field and by then sweeping the field up to 9.91 mT
at $0.81\;T_c$. The final part of the deceleration below  0.7
rad/s is shown here. As demonstrated in
Ref.~\protect\cite{Ruutu2}, during a slow deceleration
($d\Omega/dt = -1.5 \cdot 10^{-3} \;$rad/s$^{-2}$)  the total
circulation at any value of $\Omega$ corresponds roughly to that
in the equilibrium state.  The {\it inset} illustrates
schematically a transverse cross section through the rotating
container with the coaxial array configuration: the SQV lines
(small dots) are in the outer annulus and the DQV lines (large
dots) in the center.  }
    \label{EquilArray}
  \end{center}
\end{figure}

\subsection{Vortex cluster in coexistence regime}

The radial composition of a cluster with both SQV and DQV lines
can be determined by recording the amplitudes of the satellites as
a function of $\Omega$ while the rotation is slowly decelerated.
In Fig.~\ref{EquilArray} the result is shown for an array which
has been created by cooling through $T_c$ at 1 rad/s. Here the
distributions of the two vortex species are segregated: The SQV
lines annihilate first and the DQV lines later. This behavior
corresponds to the coaxial array structure, shown in the inset of
Fig.~\ref{EquilArray}, which consists of a central region of 130
DQV lines, a peripheral annulus of 140 SQV lines, and the exterior
vortex-free CF region of equilibrium width. The boundary between
the DQV and SQV regions is relatively sharp (10$\,$\% of the
radius of the array, or additionally roughly 40 DQV and 50 SQV
lines).

The coaxial configuration is stable against externally applied
perturbations. In fact, no means were found to change the
configuration, without removing vortex lines by deceleration or by
adding new ones during acceleration. The noise in the rotation
drive is of order $|\Delta \Omega| \sim 1 \cdot 10^{-3}$ rad/s.
However, no noticeable changes in the coaxial configuration were
observed when the array is maintained for several hours at
constant $\Omega$. This is not surprising since even a sinusoidal
modulation of $\Omega$ with an amplitude $\Delta \Omega \sim 0.1$
rad/s does not cause measurable changes. (The modulation by
$2\Delta \Omega$ is less than the separation between the
annihilation and critical velocity thresholds and thus the number
of vortex lines is not yet changed.) This suggests that any
spontaneous reordering of the two vortex species within the array
from the segregated domain structure to a mixed configuration is
below our measuring resolution. Thus the cool down from $T_c$ to
the measuring temperature or a magnetic field sweep after the
transition through $T_c$ are not expected to introduce changes in
the array configuration.

The segregation into two coaxial domains, with the SQV lines in
the outer annulus, is always observed when both vortex species are
found to coexist after cooling through $T_c$. We also find, by
measuring  the width of the CF annulus \cite{Ruutu2}, that  the
total number of lines adjusts itself so that the energy in
Eq.~(\ref{energy1/ab}) is minimized. This is expected, since the
total circulation corresponds to a dominant energy term. In
comparison the structure of an individual vortex line or the
spatial distribution of the two vortex species within the array,
represent much smaller energy contributions.
\begin{figure}[!t]
  \begin{center}
    \leavevmode
  \includegraphics[width=0.9\linewidth]{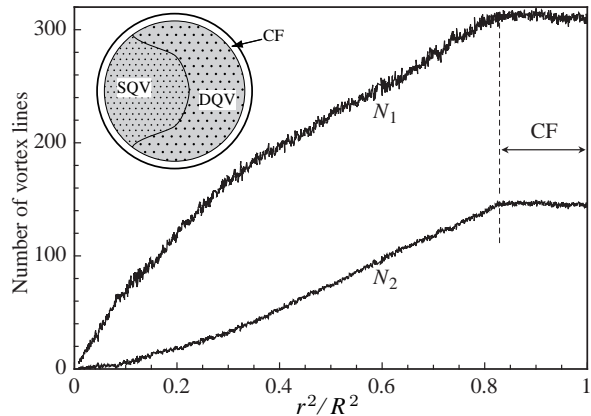}
  \bigskip
 \caption{
Radial distributions $N_1(r^2)$ and $N_2(r^2)$ in a sequentially
formed vortex cluster. The original array was an equilibrium state
of SQV lines, to which DQV lines were added later by increasing
rotation in the superfluid state. The SQV array was prepared at
29.3 bar pressure and 9.91 mT field, by cooling through $T_c$ at
$7\; \mu$K/min and 0.6 rad/s. Cooling was continued until
$0.81\;T_c$, where $\Omega$ was slowly increased to 1.4 rad/s, to
create the DQV lines. The final step was a slow deceleration to
zero, during which the radial distributions $N_1(r^2)$ and
$N_2(r^2)$ were recorded. Both $\Omega$ sweeps were performed at
$|d\Omega/dt| = 2 \cdot 10^{-3} \;$rad/s$^{-2}$. The critical CF
velocity of vortex formation at 1.4 rad/s was $\Omega_c = 0.24$
rad/s, which corresponds to a linear velocity of $v_c = 0.60$
mm/s. The {\it inset} illustrates schematically the array
configuration, a plastically deformed agglomerate of vortex lines.
      }
    \label{Glass}
  \end{center}
\end{figure}

\subsection{Sequentially formed vortex cluster}

The stability of vortex arrays in superfluid $^3$He can also be
approached from the other extreme, by studying the disorder in
artificially constructed arrays of coexisting SQV and DQV lines. These can
be prepared by making use of the metastability of existing vorticity.

We use the following procedure:  The sample is first cooled
through $T_c$ at $\Omega \leq 0.6 $ rad/s and $H > H_c$, so that
an equilibrium vortex state with only SQV lines is formed. DQV
lines are added later, by increasing $\Omega$ slowly while the
temperature is already below $T_c$, such that the desired number
of DQV lines is created at a stable critical counterflow velocity
$v_c = \Omega_c R$ \cite{Ruutu1}. In the presence of an existing
array this value of CF represents the separation between the
annihilation and critical velocity thresholds. The DQV lines are
here injected one at a  time into the existing array, presumably
from the same nucleation center on the cylindrical wall where the
superflow instability takes place (i.e. they are thought to
emanate from a rectilinear source which is close to the
cylindrical wall and parallel to the rotation axis \cite{Ruutu1}).

In Fig.~\ref{Glass} the radial decomposition of an array is shown
which has been prepared with this procedure, by adding 150 DQV
lines to an existing equilibrium array of 310 SQV lines. During
deceleration, initially both types of vortices are seen to
annihilate, but on moving closer to the center the proportion of
DQV lines is reduced and the SQV lines start to dominate. Such an
annihilation record could correspond to the configuration shown
schematically in the inset of Fig.~\ref{Glass}, which in turn
could result from the process by which the array was formed. The
DQV lines are here formed in a regular periodic process, where a
newly created DQV line arrives roughly to the same spot at the
edge of the array. It makes space for itself by pushing the
existing lines aside. In this way the DQV lines end up  residing
in a segregated pocket at the edge of the cluster, if no mixing of
lines occurs later.

From the data in Fig.~\ref{Glass} it is not possible to judge how
sharp the boundary is between the two domains and to what extent
mixing of the two vortex species might have occurred. However, no
significant change from this configuration is noticed even if the
cluster is subjected to rapid modulation of the rotation velocity
with an amplitude which does not yet add new vortices or
annihilate existing ones. On this basis we expect spontaneous
reordering or annealing not to be of importance. The sequentially
formed array is most likely a state with much reduced long-range
order, which resembles a random glassy agglomerate, but is subject
to hexagonal nearest-neighbor coordination and a circular boundary
condition.

\subsection{Homogeneity of superfluid transition}

In Chap.~\ref{Chap:Exp} we have focused on two questions: (1) the
metastability and width of the SQV $\leftrightarrow$ DQV
transition and (2) the configuration of the vortex array in the
coexistence regime. We might now simply conclude that the
segregated coaxial array, which is observed after a cool down
through $T_c$ in the coexistence regimes,  is the expected minimum
energy configuration. Since vortex clusters in $^3$He-A cannot
relax during the course of the experiments, this configuration
must also be the one which was originally formed in the transition
through $T_c$.

However, in any real experiment the superfluid transition will
acquire some degree of inhomogeneity. On some level this will lead
to  irreversibility and will interfere with the argument that the
transition is adiabatic and minimizes the total energy. Here the
main source for inhomogeneity is the thermal resistance
concentrated in the orifice of the container. The liquid $^3$He
volume  consists of the sample container and the sintered heat
exchanger in contact with the refrigerator. These are connected
together with a long tube, which ends in the orifice  on the
bottom of the sample cylinder \cite{Ruutu4}. During cool down a
thermal gradient is present, in addition to residual heat leaks,
which maintain the top end of the sample container as the warmest
point in the $^3$He column. This means that the heat flow through
the orifice is composed of the cooling rate of the liquid and the
residual heat leak. The residual heating consists of the NMR
absorption and of thermal relaxation losses seeping out of the
walls of the sample container.

At a cooling rate of 1 $\mu$K/min the heat flow from the container
is of order 0.5 nW and the thermal gradient inside the container
$\sim 0.5 \; \mu$K/mm (at $T_c \sim 2.4$ mK). The superfluid
transition emerges thus as a transition front through the orifice
on the bottom of the cylinder. Initially it adopts a hemispherical
shape \cite{VorBundle}. The width of the phase front is determined
by the local thermal gradient and thermal fluctuations.  Its
velocity can be controlled externally by adjusting the cooling
rate. The heat flow, in turn, is mainly carried by a superfluid
thermal countercurrent, which appears below $T_c$ and converges
towards the orifice. It is oriented predominantly in the axial
direction, transverse to the rotational flow.

Clearly this picture of the transition is not that of a
homogeneous second order transition (although it is probably
closer to it than in any other macroscopic system). In equilibrium
conditions at low $\Omega$ the SQV is the minimum energy vortex
structure, but thermal CF or other inhomogeneities  provide a bias
for DQV lines to be formed.  This  perturbation is concentrated in
the neighborhood of the orifice, from where vortex formation
starts during the superfluid transition. Here DQV lines are
created first and SQV lines appear later when the transition front
expands into the more homogeneous peripheral region of the
cylinder. This interpretation, which is suggested by the
dependence of the coexistence regime on the cooling rate at $T_c$
(Fig.~\ref{CoolRateDepend}) , provides a second reason for the
coaxial array configuration with the SQV lines in the outer
annulus.

According to this interpretation, the interface between the two
coaxial domains becomes a similar phase boundary as that in
Fig.~\ref{CoolRateDepend}, but now as a function of the radial
coordinate $r$ of the container, along which the cooling rate
varies as the normal-to-superfluid phase front moves through the
cylinder. This alternative explanation thus rests on the
assumption that the cooling rate at $T_c$ is one of the major
controlling factors of the phase boundary in
Fig.~\ref{CoolRateDepend}: SQV lines are formed closer to
equilibrium conditions while nonequilibrium and perturbations
favor the formation of DQV lines.


\section{ Conclusion}

By comparing experiment to numerical simulation, we conclude that
below $T_c$ in superfluid $^3$He vortex arrays are formed by
plastic deformation: Vortex lines are frozen in the array to the
configuration in which the array was formed. High energy barriers
separate different array configurations and thermally activated
processes (or any other mechanisms) are not able to anneal faults,
to create a new structure with lower energy and improved long
range order.

In arrays formed by cooling through $T_c$ in rotation, a
segregated coaxial configuration is observed, with the SQV lines
located in the outer annulus. Our numerical analysis of 2D arrays
with discrete rectilinear vortex lines confirms that this coaxial
configuration is energetically the most preferable state in the
coexistence regime. It is distinguished from all other segregated
domain structures by a gain in the outer surface energy. This
energy reduction arises on the scale of the intervortex distance
rather than the container dimensions, due to a smoother outer
interface, when the border is formed by vortex lines with a
smaller quantum number.

\acknowledgements

This work  was funded by the EU Human Capital and Mobility
Programme (grants CHGECT94-0069 and ERBFMGECT980122) and by INTAS
(grant 96-0610). NBK acknowledges support from the Russian
Foundation for Basic Research (grant 99-02-16043).

\end{document}